\documentclass{desyproc}

\providecommand{\pp}{$p$-$p$}
\providecommand{\sqrts}{\sqrt{s}}
\providecommand{\PYTHIA}{{\sc pythia }}
\providecommand{\HERWIG}{{\sc herwig }}

\def\mean#1{\ensuremath{\left<#1\right>}}

\begin{document}


\title{Forward jets physics in ATLAS, CMS and LHCb}

\author{{\slshape David d'Enterria$^1$}\\[1ex]
$^1$ICC-UB \& ICREA, Univ. de Barcelona, 08028 Barcelona, Catalonia}

\contribID{denterria\_david}

\desyproc{DESY-PROC-2009-xx}
\acronym{EDS'09} 
\doi  

\maketitle

\begin{abstract}
The capabilities of the ATLAS, CMS and LHCb detectors to reconstruct jets at
forward rapidities ($|\eta|>$~3) in \pp\ collisions at the CERN Large Hadron Collider are 
reviewed. The QCD and Higgs physics motivations for such measurements are summarised.
Details are given on 
studies that provide information on the parton structure and evolution
at small values of fractional momenta in the proton. 
\end{abstract}


\section{Introduction}

\begin{figure}[htb]
\centering
\includegraphics[width=0.48\textwidth,height=6.cm]{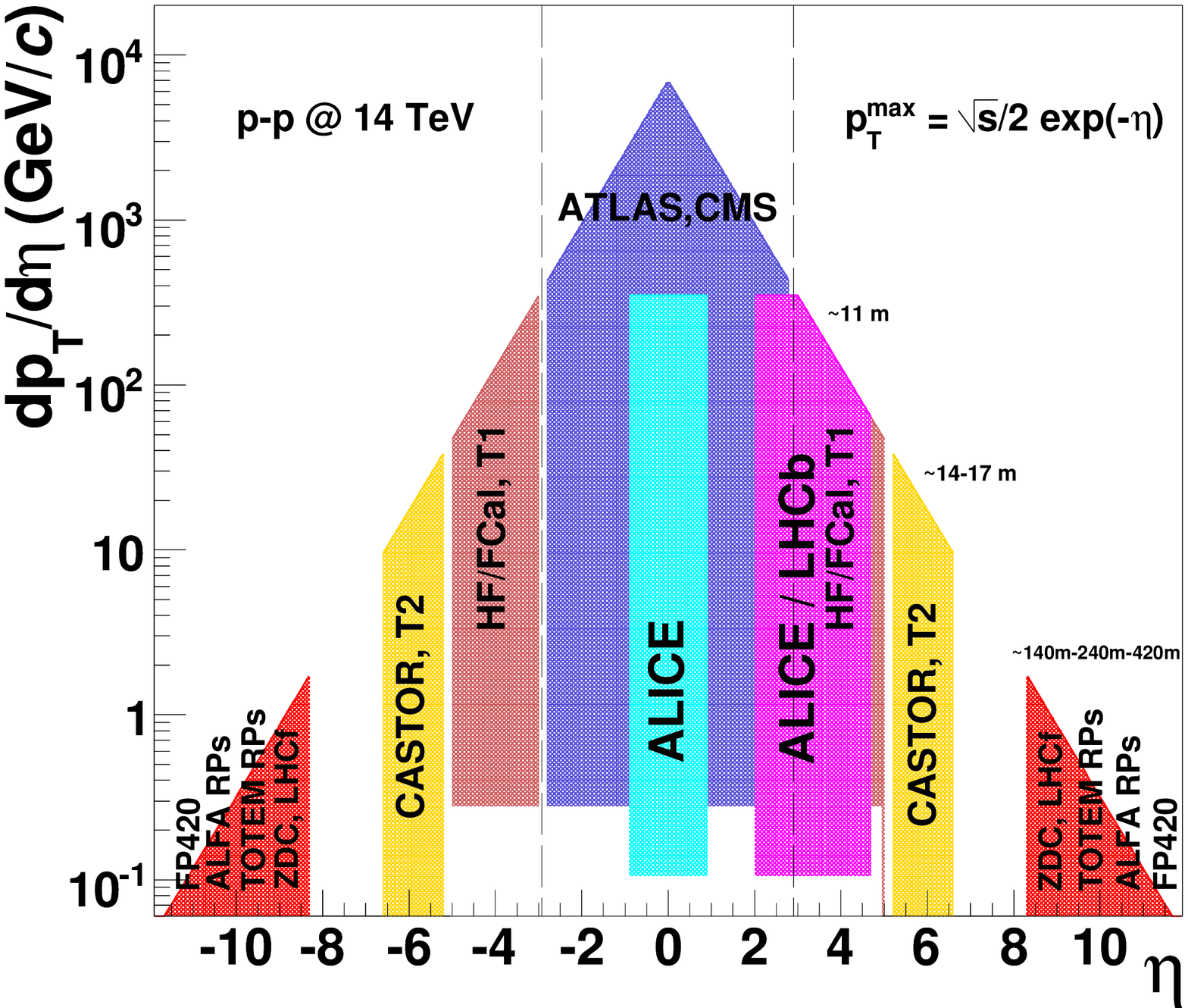}\hspace{0.2cm}
\includegraphics[width=0.49\textwidth,height=6.1cm]{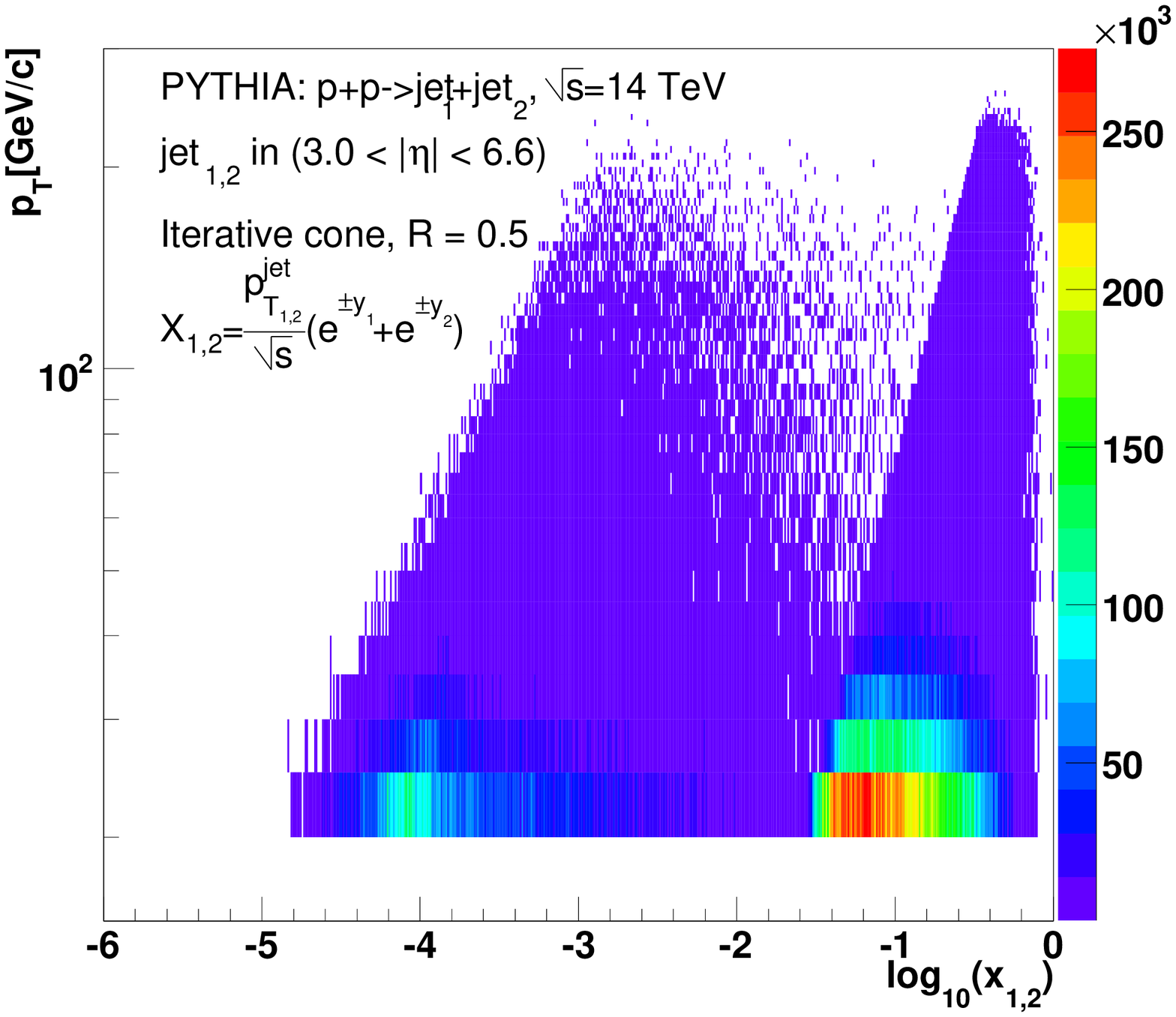}
\caption{Left: Acceptance of the LHC detectors in the $p_T$--$\eta$ plane ('forward' detectors are beyond
the dashed vertical line)~\protect\cite{dde_fwdlhc}.
Right: Log($x_{1,2}$) distribution of two partons producing at least one jet above $p_T$ = 20~GeV/c
at forward rapidities in \pp\ collisions at $\sqrts$~=~14~TeV~\protect\cite{scerci_dde}. 
\label{fig:1}}
\end{figure}

The ATLAS, CMS and LHCb experiments\footnote{ALICE has jet reconstruction capabilities at central~\cite{alice_jets} 
but not forward rapidities.} feature detection capabilities at forward rapidities 
($|\eta|>$~3, see Fig.~\ref{fig:1} left) which allows them to reconstruct jets in a kinematic 
range of interest for various Higgs and QCD physics studies in \pp\ collisions at TeV energies.\\

On the one hand, having the possibility to reconstruct jets beyond $|\eta|\approx$~3 in ATLAS 
and CMS is crucial  to signal Higgs boson production in vector-boson-fusion (VBF) processes, 
$qq \xrightarrow{VV} qqH$ (with $V=W,Z$), where the valence quarks $q$ from each proton
fragment into jets in the forward and backward hemispheres~\cite{Djouadi:2005gi}.
The presence of such low-angle jets 
is instrumental to significantly reduce the QCD backgrounds in various VBF Higgs discovery 
channels at the LHC, particularly for low $H$ masses~\cite{CMS_TDR2,atlas_asai}. 
In the case of LHCb, given the excellent secondary vertex capabilities of the detector,
forward jet studies have focused on the reconstruction of $b$-jets aiming at the $H \to b \bar b$ 
decay channel in Higgs production associated with vector bosons (about one third of the
cross section falls within the LHCb acceptance)~\cite{lhcb_jets}.\\

On the other hand, forward jet production is in its own right an interesting perturbative QCD (pQCD) process whose 
study yields important information on the underlying parton structure and its dynamical evolution in the proton.
In particular, it provides valuable information on the gluon density $xG(x,Q^2)$ in a regime of low 
momentum fraction, $x$~=~$p_{\mbox{\tiny{\it parton}}}/p_{\mbox{\tiny{\it hadron}}}<$~10$^{-2}$, 
where standard deep-inelastic $e$-$p$ data can only indirectly constrain its value~\cite{dde_lowx},
and where its evolution is expected to be affected by non-linear QCD dynamics~\cite{cgc}.
Indeed, in \pp\ collisions, the {\it minimum} parton momentum fractions probed in each proton in 
a  $2\rightarrow 2$ process with a jet of momentum $p_T$ produced at pseudo-rapidity $\eta$ are
\begin{equation}
x_{2}^{min} = \frac{x_T\,e^{-\eta}}{2-x_T\,e^{\eta}}\;\;,\;\;\mbox{ and }\;\; x_{1}^{min} = \frac{x_2\,x_T\,e^{\eta}}{2x_2-x_T\,e^{-\eta}}\;\;,
\mbox{ where } \;\; x_T=2p_T/\sqrt{s}\,,
\label{eq:x2_min}
\end{equation}
i.e. $x_2^{min}$ decreases by a factor of $\sim$10 every 2 units of rapidity. The extra $e^\eta$ 
lever-arm motivates the interest of {\it forward} jet production measurements to study the PDFs
at small values of $x$. From Eq.~(\ref{eq:x2_min}), it follows that the measurement at the LHC 
of jets with transverse momentum $p_{T}$~=~20~GeV/c at rapidities $\eta\approx$~5 allows 
one to probe $x$ values as low as $x_{2}\approx 10^{-5}$ in partonic collisions with highly 
asymmetric longitudinal momenta in the initial-state (Fig.~\ref{fig:1}, right).\\

In this contribution we summarise first the forward jet reconstruction capabilities 
of the three LHC experiments (Section~\ref{sec:exp}) and, then, in 
Sections~\ref{sec:spec} and~\ref{sec:MN} we present simulation studies
of two CMS forward-jet measurements~\cite{scerci_dde}:
\begin{enumerate}
\item single inclusive jet cross section at moderate transverse momenta ($p_{T}\approx$ ~20~--~120~GeV/c),
\item 
azimuthal (de)correlations of ``Mueller-Navelet''~\cite{mueller_navelet} dijet events, characterised by 
jets with similar $p_T$ separated by a large rapidity interval ($\Delta\eta\approx$~6~--~10),
\end{enumerate}
which are 
sensitive, respectively, to the small-$x_2$ (and high-$x_1$) proton PDFs, as well as to low-$x$ QCD 
evolution of the BFKL~\cite{bfkl}, CCFM~\cite{cascade} and/or saturation~\cite{cgc} types.

\section{Experimental performances}
\label{sec:exp}

\begin{figure}[htb]
\includegraphics[width=0.40\textwidth,height=5.8cm]{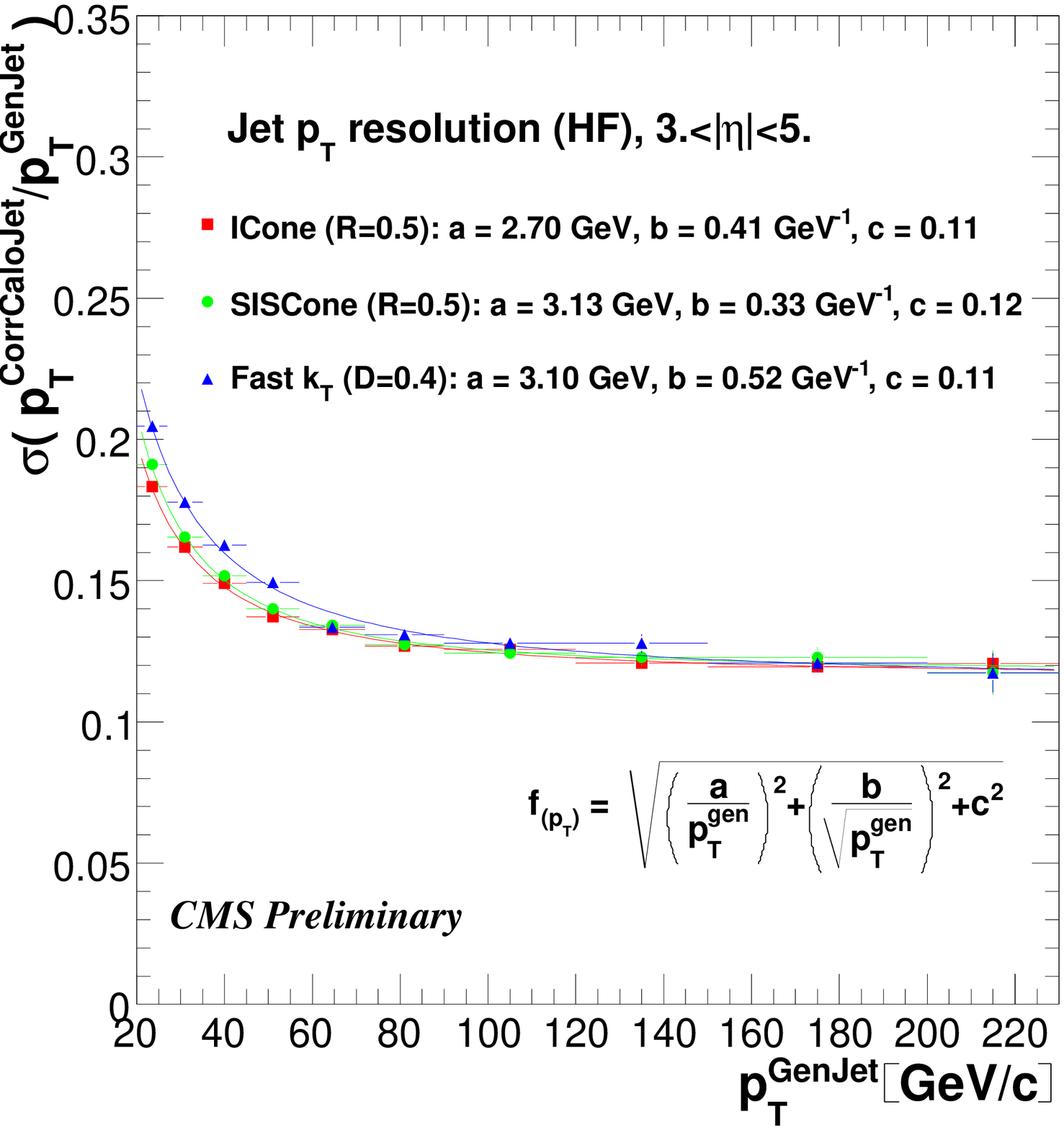}
\includegraphics[width=0.60\textwidth,height=5.56cm]{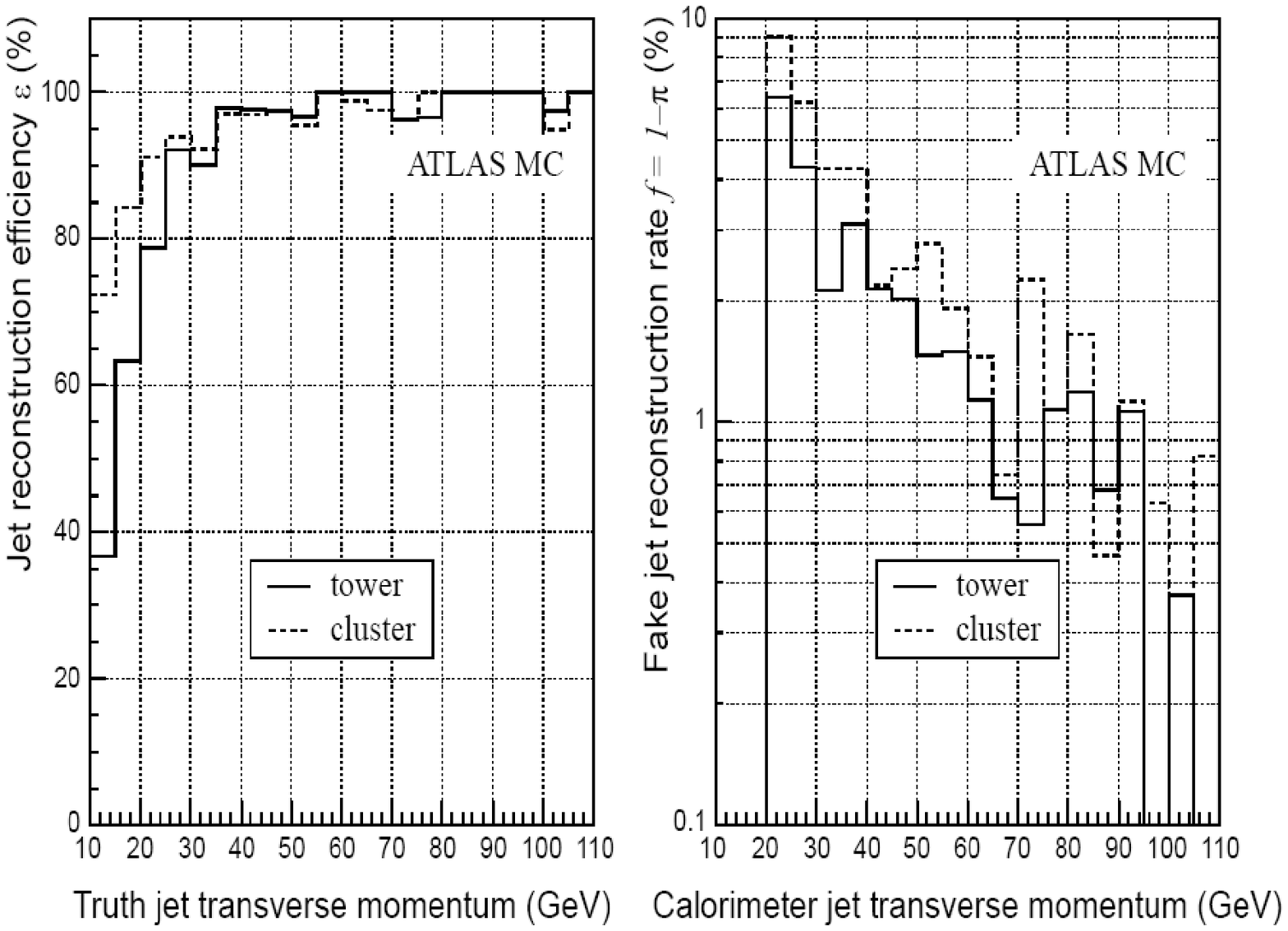}
\caption{
Left: Forward jet relative $p_T$ resolutions for various jet algorithms in the CMS HF calorimeter~\cite{scerci_dde}. 
Right: Forward jet reconstruction efficiency as function of the true jet $p_T$, and fake reconstruction
rate versus the reconstructed jet $p_T$ in the ATLAS FCal calorimeter~\cite{atlas_tdr}.
\label{fig:2}}
\end{figure}

In ATLAS and CMS, jets can be reconstructed calorimetrically at forward rapidities in the 
FCal~\cite{fcal} and HF~\cite{hf} calorimeters\footnote{In addition, in CMS one can further extend 
jet reconstruction up to $|\eta|\approx$~6.6 with the CASTOR detector~\cite{castor}.} 
(3$<|\eta|<$5), by means of standard jet algorithms of the cone or sequential-clustering 
types~\cite{jets_houches}. The jet radii are often chosen relatively small (e.g. ${\cal R}=0.5$ 
for the cone and $D=0.4$ for the $k_T$ algorithms) so as to minimise the effects of hadronic 
activity inside the jet due to the underlying event and beam-remnants.
Figure~\ref{fig:2} (left) shows the energy resolution for forward jets 
reconstructed in CMS with three different algorithms (iterative cone,
SISCone  
and $k_T$)~\cite{scerci_dde}. 
The obtained $p_T$ relative resolutions are of $\mathcal{O}(20\%)$ at 20 GeV/c decreasing to 
10\% above 100 GeV/c. Similar results are obtained for ATLAS~\cite{atlas_tdr}. 
We note that though the forward calorimeters have a coarser granularity than the  barrel and endcap ones, 
the energy resolution is {\it better} in the forward direction than at central rapidities because (i) 
the {\it total} energy of the jet is boosted at forward rapidities, and (ii) the forward jets are more 
collimated and, thus, the ratio of jet-size/detector-granularity is more favourable.
The position ($\eta$, $\phi$) resolutions (not shown here) for forward jets are also very good: 
$\sigma_{\phi,\eta}~\approx$~0.045 at $p_T$~=~20 GeV/c, improving to $\sigma_{\phi,\eta}\approx$~0.02 
above 100 GeV/c~\cite{scerci_dde}. Good $\phi$-$\eta$ resolutions are important when it comes 
to detailed studies of the azimuthal decorrelation as a function of the pseudorapidity separation
in events with forward-backward dijets (see Section~\ref{sec:MN}).
Figure~\ref{fig:2} (right) shows the efficiency and purity of forward jets reconstructed 
with the seeded cone finder (${\cal R}=0.4$) in the ATLAS FCal calorimeter~\cite{atlas_tdr}.
Above $\sim$35~GeV/c, the efficiency saturates at around 95\% with a purity below 4\%. 

In LHCb, jet reconstruction has focused on $b$-jets given the excellent vertexing
capabilities of the detector. The physics motivation is so far centered on the measurement of 
the $H \to b \bar b$ channel for intermediate-mass Higgs production associated with vector 
bosons (30\% of such a signal falls within the LHCb acceptance)~\cite{lhcb_jets}. Both 
seeded-cone- and $k_T$-algorithms have been tested including information from the calorimeters 
and the tracking devices in a ``particle flow'' type of approach. A neural-network is trained to 
identify $b$-jets and optimise the jet energy reconstruction. The main current limitation of the 
measure is the saturation of the calorimeters (designed originally mostly for single particle triggering/measurements 
at low and moderate $p_T$'s) for jets with total energy beyond $E_{tot}\approx$~1.5~TeV 
(i.e. $p_T\approx$~20~--~150~GeV/c for $\eta\approx$~3~--~5) given the very strong 
$\cosh(\eta)$ total-momentum boost at large rapidities.

\section{Inclusive forward jet spectrum: Low-$x$ PDFs}
\label{sec:spec}

\begin{figure}[htbp]
\includegraphics[width=0.49\textwidth,height=6.1cm]{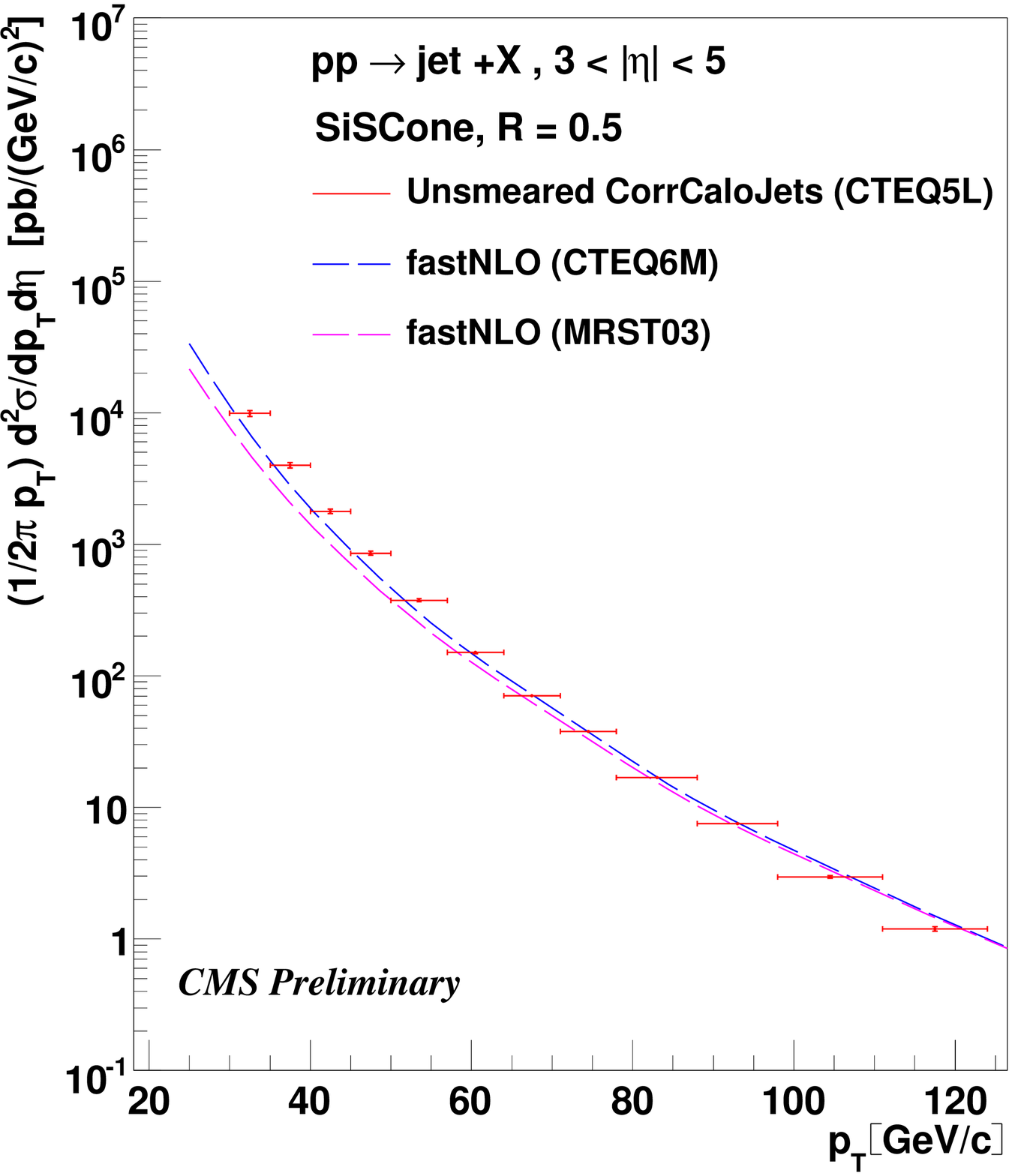}
\includegraphics[width=0.5\textwidth,height=6.cm]{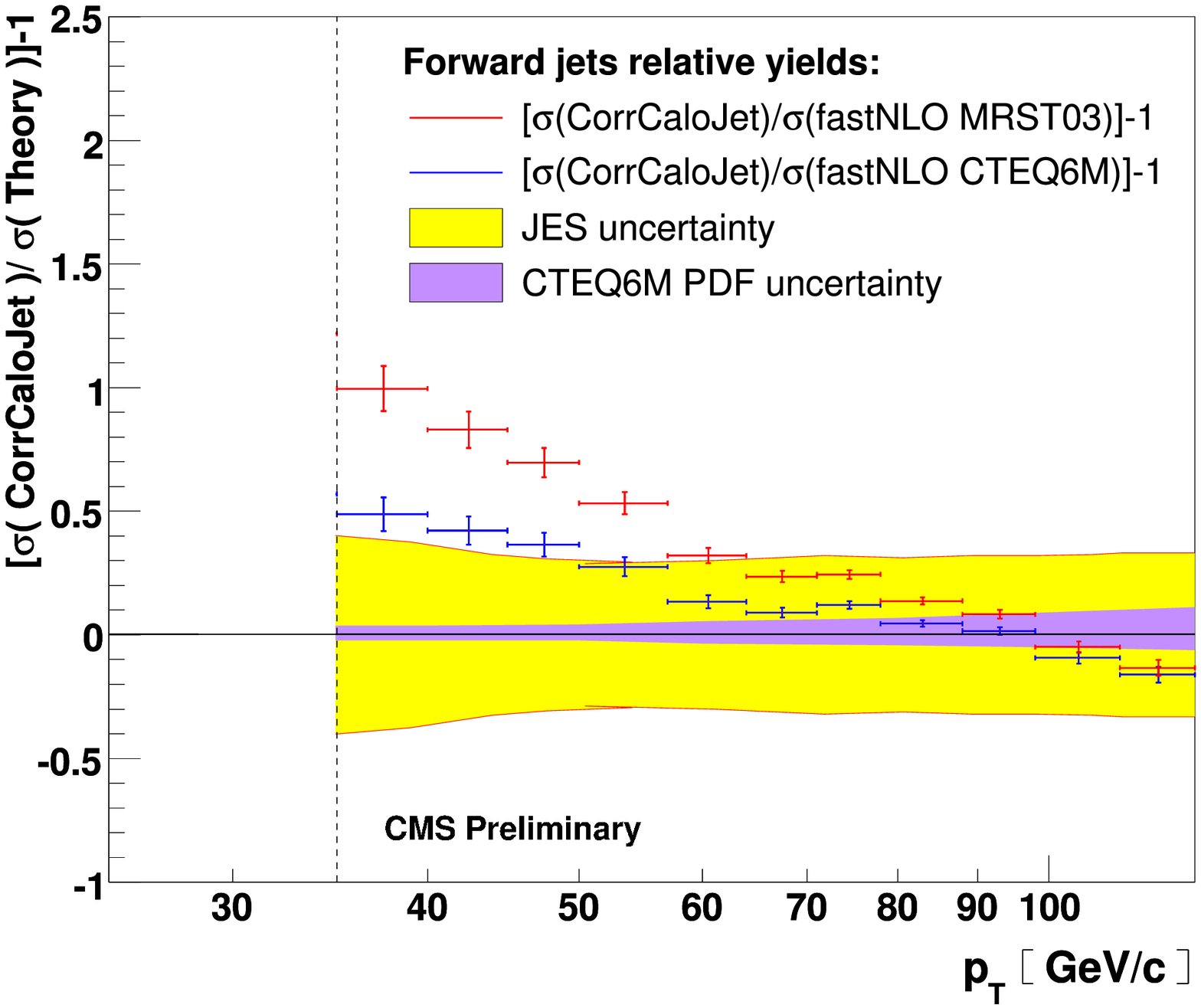}
\caption{Left: Reconstructed forward jet spectrum (only stat errors shown) in \pp\ at 14 TeV compared to fastNLO predictions 
with MRST03 and CTEQ6.1M PDFs. Right: Percent differences between the forward jet $p_T$ spectrum 
and the two fastNLO predictions. The yellow band shows the propagated yield uncertainty for a 
5~--~10\% jet-energy scale (JES) error.}
\label{fig:3}
\end{figure}

Figure~\ref{fig:3} (left) shows the forward jet $p_T$ spectrum generated with 
\PYTHIA and reconstructed in the CMS HF calorimeter with the SISCone finder~\protect\cite{scerci_dde}.
The spectrum is compared to fastNLO jet predictions~\cite{fastnlo} with the MRST03 and 
CTEQ6.1M PDFs. The right plot shows the percent differences between the reconstructed spectrum and the
two theoretical predictions. The single jet spectra obtained for different PDFs are similar at high $p_{T}$, 
while differences as large as $\mathcal{O}(60\%)$ appear below $\sim$60~GeV/c. 
The error bars include the statistical (a total integrated luminosity of 1~pb$^{-1}$ is assumed) and 
the energy-resolution smearing errors. The thin violet band around zero is the PDF uncertainty from the CTEQ6.1M set alone.
The main source of systematic uncertainty is due to the calibration of the jet energy-scale (JES). 
Assuming a conservative 5~--~10\% JES error, one finds propagated uncertainties of the order 30~--~40\% in the jet 
yields at $p_T=$~35~--~60~GeV/c (yellow band) which are similar to the theoretical uncertainty associated 
to the PDF choice. If the JES can be improved at the 5\% level or below, and the PDF uncertainties 
are indeed as large as the differences between MRST03 and CTEQ6M, a forward jet measurement could help 
constrain the underlying PDF in global-fit analyses.

\section{Forward-backward dijet correlations: Low-$x$ QCD}
\label{sec:MN}

\begin{figure}[htb]
\includegraphics[width=0.49\textwidth,height=5.7cm]{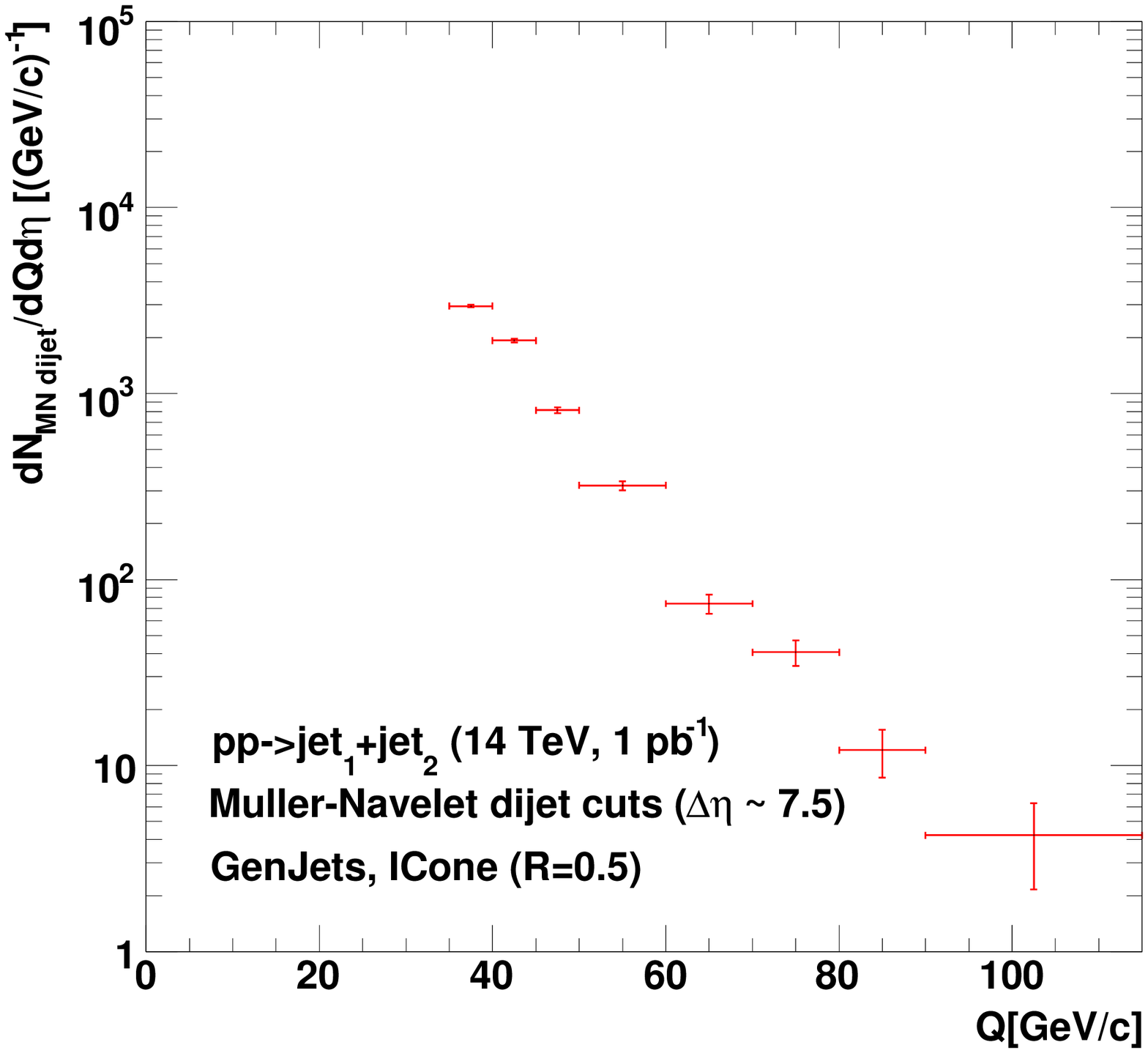}
\includegraphics[width=0.50\textwidth,height=5.7cm]{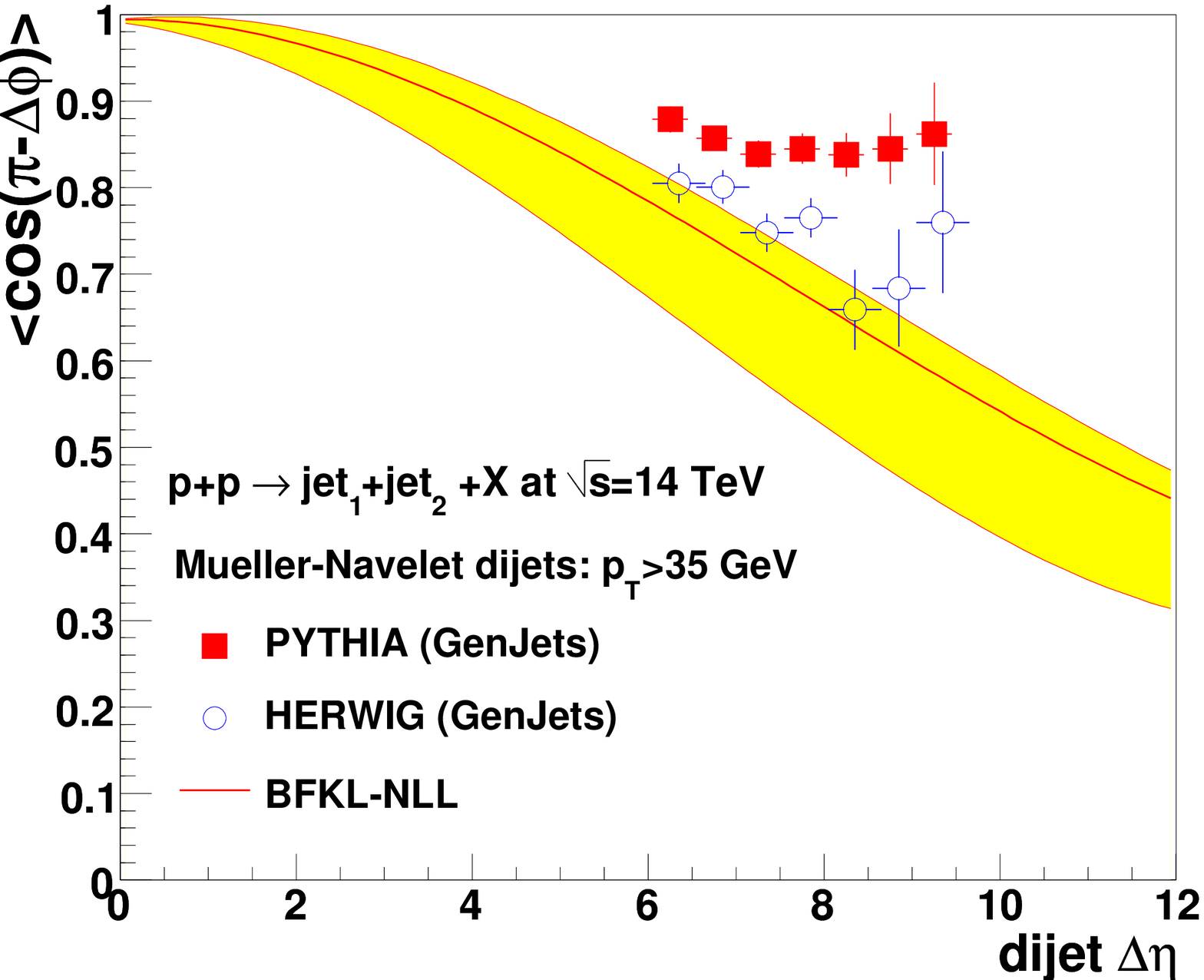}
\caption{CMS study of dijet events passing the Mueller-Navelet cuts (see text)~\cite{scerci_dde}. 
Left: Expected \PYTHIA yields (1~pb$^{-1}$) for $\Delta\eta \approx$~7.5. 
Right: Average of $cos(\pi - \Delta \phi)$ versus $\Delta \eta$ in {\sc pythia}-, {\sc herwig}-generated events
compared to BFKL (yellow band)~\cite{sabiovera} analytical estimates.
\label{fig:muller_navelet_jets}}
\end{figure}

The interest in forward jet measurements goes beyond the {\it single} inclusive cross sections: 
the production of {\it dijets} with similar $p_T$ but separated by large rapidities, the so-called 
``Mueller-Navelet jets''~\cite{mueller_navelet}, is a particularly sensitive measure of non-DGLAP 
QCD evolutions. The large rapidity interval between the jets (e.g. up to $\Delta\eta \approx$~12 
in the extremes of CMS forward calorimeters) enhances large logarithms of the type 
$\Delta\eta \sim log(s/p_{T,1}p_{T,2})$ which can be appropriately resummed within the 
BFKL~\cite{sabiovera}, CCFM~\cite{cascade} and/or saturation~\cite{marquet,iancu08}
frameworks. One of the phenomenological implications of this type of dynamics
is an enhanced radiation between the two jets which results in a larger azimuthal 
decorrelation for increasing $\Delta\eta$ separations compared to collinear pQCD approaches. 
CMS~\cite{scerci_dde} has carried out an analysis with~\PYTHIA\cite{pythia6.4} and \HERWIG\cite{herwig6}
selecting events with forward jets (ICone, ${\cal R} =$~0.5) which satisfy the following Mueller-Navelet 
(MN) type cuts:
\begin{itemize}
\item $p_{T,i} > 35$~GeV/c (good parton-jet matching and good jet trigger efficiencies in HF)
\item $|p_{T,\,1} - p_{T,\,2}| < 5$~GeV/c (similar $p_{T}$ to minimise DGLAP evolution)
\item $3 <|\eta_{1,2}|< 5$ (both jets in HF) 
\item $\eta_1 \cdot \eta_2 < 0$  (each jet in a different HF, i.e. their separation is $\Delta\eta\gtrsim$~6)
\end{itemize}
The data passing the MN-cuts are divided into 4 equidistant pseudorapidity bins 
with separations $\Delta\eta$=6.5, 7.5, 8.5 and 9.5 and the dijet cross section  
computed as $d^2\sigma/d\eta dQ = N_{jets}/(\Delta\eta \Delta Q \int\mathcal{L}\mbox{dt})$, 
where $Q = p_{T,1} \approx p_{T,2}$ and  $N$ is the observed number of  jets in the $\Delta\eta, \Delta Q$ 
bin. For 1~pb$^{-1}$, 
one expects a few 1000s (100s) MN jets with separations $\Delta\eta>$~6~(9). Figure~\ref{fig:muller_navelet_jets}, 
left, 
shows the expected \PYTHIA yields passing the MN cuts for $\Delta\eta\approx$~7.5.
The obtained dijet sample appears large enough to carry out detailed studies of the $\Delta\eta$
dependence of the yields, and look e.g. for a possible ``geometric scaling'' behaviour in the Mueller-Navelet yields~\cite{iancu08}.\\

An enhanced azimuthal decorrelation for increasing rapidity separation, measured e.g. by the average value 
(over events) of the cosine of the $\Delta\phi$ difference between the MN jets $\mean{cos\,(\pi - \Delta \phi)}$ 
versus the $\Delta\eta$ between them, is the classical ``smoking-gun'' of BFKL radiation~\cite{sabiovera,marquet}.
One expects $\mean{cos(\pi - \Delta \phi)}$ = 1 (0) for perfect (de)correlation between the two jets.
The results are shown in Fig.~\ref{fig:muller_navelet_jets} (right)  
for the two highest-$p_T$ jets in the event passing the MN cuts. Only the dominant (statistical) errors are presented. 
At the Monte Carlo truth level (not shown here), the originating partons in \PYTHIA or \HERWIG are almost exactly back-to-back 
for all $\Delta\eta$ in each such jet-pair events. Yet, at the {\it generator-level}, the $\mean{cos(\pi - \Delta \phi)}$
decorrelation increases to 15\% (25\%) for \PYTHIA ({\sc herwig}), 
$\mean{cos(\pi - \Delta \phi)}\approx$~0.85~(0.75), 
due to parton showering and hadronization effects. 
Yet, the forward dijet decorrelation observed in both MCs is smaller (and less steep as a function of $\Delta\eta$) 
than found in BFKL approaches (yellow band)~\cite{sabiovera,marquet}.


\section*{Summary}

We have summarised the forward jet reconstruction capabilities of the ATLAS, CMS and LHCb experiments.
The measurement of forward jets opens up the possibility to carry out interesting studies in the 
Higgs (tagging of vector-boson-fusion or Higgs-to-$b\bar b$ in associated $W\,Z$ production)
and QCD (low-$x$ parton densities and dynamics) sectors of the Standard Model.


\section*{Acknowledgments}

The author thanks Salim Cerci for his valuable collaboration with the CMS forward jets analysis,
as well as Marco Musy for discussions on LHCb jet reconstruction. 
Support by the 7th EU Framework Programme (contract FP7-ERG-2008-235071) is acknowledged.


\begin{footnotesize}

\end{footnotesize}


\end{document}